\documentclass[12pt]{article}
\title{Running coupling constant of background perturbation\\
theory and lattice  interquark interaction.}
\author{A.M.Badalian\\
Institute for Theoretical and Experimental Physics,\\
B.Cheremushkinskaya 25, 117259 Moscow, Russia}
\date{}
\begin{document}
\maketitle

\newcommand{\be}{\begin{equation}}
\newcommand{\ee}{\end{equation}}

\def\la{\mathrel{\mathpalette\fun <}}
\def\ga{\mathrel{\mathpalette\fun >}}
\def\fun#1#2{\lower3.6pt\vbox{\baselineskip0pt\lineskip.9pt
\ialign{$\mathsurround=0pt#1\hfil##\hfil$\crcr#2\crcr\sim\crcr}}}

\centerline{\bf{Abstract}}

\vspace{5mm}
The conventional parametrizations of lattice static interquark force are
shown to produce a mismatch of coupling constant near the matching point
$R_m \approx 0.2$fm. The running coupling constant of the background
perturbation theory yields instead a selfconsistent description of lattice
data over all interval of distances, $0.05$fm$ \la R \la$1.1.fm, and is an
analytic function with the property of asymptotic freedom at small $R$ and
it is freezing at large $R$.

\newpage

\begin{center}
{\Large Running coupling constant of background perturbation}

{\Large theory and lattice  interquark interaction.}

\vspace{1cm}
{\Large A.M.Badalian}

\vspace{5mm}
{\large Institute for Theoretical and Experimental Physics,}

{\large B.Cheremushkinskaya 25, 117259 Moscow, Russia}

\end{center}

\bigskip

1. Recently the interaction betwen static quark and antiquark was
intensively studied in lattice version of pure gauge $SU(3)$ and $SU(2)$
theories [1-6]. These investigations seem to be very important because
just in lattice approach both perturbative (P)  and nonperturbative (NP)
effects and their interference can be measured and thus we have an
opportunity to get an information out of the framework of pure P theory. To
this end static potential $V(R)$, or force $F(R)=\frac{dV}{dR}$, has to be
measured over large interval of interquark separations R which is possible
only on large volume  lattices. Such lattice data were presented in
\cite{4,5} for $SU(3)$ gauge theory  and here we shall mostly discuss
results obtained in \cite{4} (lattice $36^4$ at $\beta=6.5$, $a^{-1}=4,13$
GeV or lattice spacing $a \cong 0.048$ fm) where the force was measured on
the large interval $a \la R \leq 23a$ or $0.05$ fm $\la R \la 1,1$ fm.

At small distances, $R \la $0.2 fm, the regime of asymptotic freedom was
observed and lattice data were well described by only $P$  contribution
\cite{4,5}

\be
F^P(R) = \frac{4}{3}~\frac{\alpha_F(R)}{R^2};
\ee
with the effective running coupling constant (c.c.) $\alpha_F(R)$ defined by
two-loop expression which was taken in \cite{4,5} in the form:

\be
\alpha_F(R) = \frac{4\pi}{b_0 t}~\Biggl [ 1 +
\frac{b_1}{b^2_0}~\frac{ln~t}{t}\Biggr ]^{-1}, ~~~~t= ln
~\frac{1}{(\Lambda_R R)^2} \cdot
\ee
Here $b_0 = 11$  and $b_1 = 102$  are the usual coefficients for the
$\beta$-function with the number of flavours $n_f=0$.

The "experimental"  lattice points $\alpha(R)$ were defined in following way

\be
\alpha\Biggl ( \frac{1}{2}(R_1 + R_2)\Biggr ) = \frac{3}{4}R_1 R_2 ~
\frac{V_c (R_1) - V_c(R_2)}{R_1 - R_2},
\ee
where $V_c(R)$ is the corrected static potential which takes into account
the lattice artefacts responsible for the lack of rotational invariance. If
we accept that NP contribution is small at small separations, $R < 4a$,
then $\alpha(R)$ (3) has to coincide with $\alpha_F(R)$ in (1). Below we
give values of $\alpha(R)$  at two points $R/a$   which are important for
our further discussion (the numbers are taken from Table 3 in paper
\cite{4}):

$$ \frac{R}{a} = 3.3839  ~~~\mbox{or}~~~ R\cong 0.16~fm, ~~~\alpha(R) =
0.317~~~(10)~(2)$$

\be
\frac{R}{a} = 3.9241 ~~~\mbox{or}~~~ R\cong 0.19 fm, ~~~\alpha(R) = 0.333
~~~(6)~(1)
\ee
From the lattice analysis \cite{4} it is clear that these two
points $R/a$ lie just near the boundary of the region (denote it as $R_m = 4a
\cong 0.191$ fm) where P contribution (1) dominates. At $R \ga 4a$, when
both P and NP contributions are becoming important, another parametrization
is widely used [4-6], it corresponds to the Cornell potential type of
interaction.  This force  can be written (in lattice units)
as

\be
F(R)   = a^2\frac{\Delta V}{\Delta R} = \frac{E}{(R/a)^2} + Ka^2, ~~~R \ga
4a
\ee
and from lattice fit the following values of $K$  and $E$  were obtained
$\Biggl ( \frac{\chi^2}{N_{d.o.f.}} = \frac{9.1}{13}\Biggr ) $

\be
E = 0.278 ~(7), ~~~~~Ka^2 = 0.0114~(2)
\ee
The string tension $K$ in (5)  can be found from (6) and for $a^{-1} =
4,13$GeV it gives $\sqrt{K} = 440$MeV. Also from (6) it follows that the
asymptotic value of c.c. is

\be
\alpha_F(asym) = \frac{3}{4}E = 0.2085~(5),
\ee
which we can  compare with c.c. values (4) near the matching point $R_m=4a$.
  From this comparison one can see that the asymptotic value (7) is
about 40\% less than c.c. $\alpha(R_m)\cong 0.33$  which was directly
measured on the lattice. Thus if  the parametrization (5) is used at $R \ga
4a$ and the purely P force (1) is taken at $R < 4a$ then the resulting
function $\alpha_F(R)$ is discontinuous.

To overcome this difficulty and get selfcosistent description of static
lattice potential or  force we suggest to use another approach
which takes into account the behaviour of running c.c. in background vacuum
fields. As it was shown by Yu.Simonov in \cite{7,8} this running c.c.,
$\alpha_B(R)$,  is an analytical function at all distances $R$  and
its expression  simplifies at small and large distances.
At small distances, $R^{-2} \gg \Lambda^2_R$, $\alpha_B(R)$ coincides with
standard $P$ coupling constant, i.e. manifests the property of asymptotic
freedom.  At large values of $t_B$ (see formula (9)) the following
expression for $\alpha_B(R)$ was obtained,

\be
\tilde{\alpha}_B(R)  = \frac{4\pi}{b_0 t_B} \Biggl ( 1 - \frac{b_1}{b^2_0}~
\frac{ln~t_B}{t_B}\Biggr ), ~~~R m_B  \gg 1
\ee
Here

\be
t_B = ln ~\frac{1+R^2 m^2_B}{\Lambda^2_R R^2}
\ee
and $m_B$ is a characteristic background mass which  defines the asymptotic
(exponential) falling off of self-energy part of two-point function
$\prod(x,y)$. This mass can depend on the  process or channel considered
and, in particular, in $e^+e^-$ annihilation $m_B$  coincides with the
lowest hybrid mass, $M_h(0^{++})$\cite{8}. The lattice measurements \cite{9}
has given $M_h(0^{++}) \cong 1.5$GeV.

In this letter we analyze $\alpha_B(R)$ as obtained from static
$Q\bar{Q}$ force. On the lattice the  static interaction
was measured for Eucledean  time $T_E$ which is
usually not large, e.g.  in \cite{4} lattice measurements were done at $T_E
\cong 3a \div 5a \sim 0.2$fm and for such $T_E$  quark and antiquark are
situated rather close to each other.  In this case background mass $m_B$ can
be smaller $M_h$ but larger than two  constituent gluon masses $M(2g) =
2Mg \approx 1.0$GeV \cite{8}.  In our  analysis we shall use two
different values of $m_B$:

\be
m^{(1)}_B = M_h(0^{++}) = 1.5 GeV,~~~m^{(2)}_B \cong M(2g) = 1.0 GeV
\ee

From (8)  one can see that this expression,  derived at large
$t_B$,  has a correct $P$ behaviour at small $R$, i.e. has the
property of asymptotic freedom.  The important second feature of
$\alpha_B(R)$ is that it has no infrared singularity and can be applied at
any distances.  Also the expression (8) gives the correct limit at large
$R$ when $\alpha_B(R)$ tends to a constant value as it was shown in
\cite{7}. For this reasons we can use (8) as a convenient interpolation
formula over all interval of interquark separations $R$.

Now to coordinate our analysis with lattice approach   we
introduce,  instead of (8), slightly modified expression for $\alpha_B(R)$,

\be
\alpha_B(R)  = \frac{4\pi}{b_0 t_B}~\Biggl ( 1 +
\frac{b_1}{b^2_0}~\frac{ln~t_B}{t_B}\Biggr )^{-1},
\ee
which is going over into the $P$ expression (2) at $m_B = 0$. In (11) $t_B$
is  defined by (9). From (11) one can easily see that $\alpha_B(R)$  is
freezing at  large distances, $\alpha_B(R)\to const$  because $t_B \to
const$ if $m^2_B R^2 \gg 1$:

\be
t_B (R \to \infty)  \longrightarrow   t_{\infty} =
ln~\frac{m^2_B}{\Lambda^2_R}\cdot
\ee
The asymptotic value of $\alpha_B(R)$  is given by

\be
\alpha_B(asym) = \frac{4\pi}{b_0 t_{\infty}}\Biggl ( 1 + \frac{b_1}{b^2_0}~
\frac{ln~t_{\infty}}{t_{\infty}} \Biggr )^{-1}
\ee
To find $\alpha_B$(asym) we should know, besides $m_B(10)$, also
$\Lambda_R$. Here  we shall choose $\Lambda_R$   to be very close
to those values $\Lambda_R$ which were obtained in \cite{4},
$\Lambda_{\overline{MS}} = 256(20)$MeV, and in \cite{5},
$\Lambda_{\overline{MS}}$ = 293 (18) ${+25\choose -63}$ MeV.  As it is well
known \cite{10}  $\Lambda_R = 1.048 \Lambda_{\overline{MS}}$ i.e.
$\Lambda_R$  defining the force and $\Lambda_{\overline{MS}}$ practically
coincide within lattice errors.

Then in the presence of background fields the force $F(R)$ can be presented
as a sum

$$F(R) = F^{(PB)}(R) + F^{NP}(R), \eqno{(14a)}$$
where instead of pure $P$  term (1) we introduce the
 modified  term,

$$F^{PB} (R) = \frac{4}{3}~\frac{\alpha_B(R)}{R^2}~~~~~\eqno{(14b)}$$
with $\alpha_B(R)$  taken  in the form (11).

Some remarks are needed about our choice of NP part of interaction.
We shall consider two possibilities: \underline{case A}, when NP
potential is defined by linear potential at all distances or $F^{NP}(R) =
Ka^2 = const$ as in (5), and \underline{case B}, when NP interaction is
defined through bilocal correlators $D(x)$ with a finite correlation length
$T_g$ \cite{11}.  The first measurements of correlators and $T_g$ were done
in \cite{12} where the value $T_g \cong 0.2$fm (or $\delta = T_g^{-1} \cong
1$GeV)  was found.  Note that the case A corresponds to vanishing
correlation length $T_g$, or $\delta \gg 1$Gev.  The potential and
force, corresponding to a finite $T_g$ and the exponential form of the
correlator $D(x)$  were calculated in \cite{13},

$$
F^{NP}(R)  = a^2\frac{d\varepsilon}{dR} =
2a^2\int\limits^{\infty}_0~dv~\int\limits^R_0~d\lambda~D(\sqrt{\lambda^2
+ v^2})= \frac{2}{\pi}K a^2 \int^{R\delta}_0 ~tK_1(t) dt \eqno{(15)} $$
where $K_1(t)$ is the modified Bessel function.  From (15) one can see that
(i) if $T_g$   is finite then the force is diminishing $(F(R) \sim R)$ at $R
\to 0$;  (ii) $F(R)$ is approaching the constant value, $Ka^2$, only at
distances $R_{form} \ga 3T_g \sim 0.6$fm.  This  second statement is in
agreement with Bali, Schilling result \cite{14} that the formation of string
takes place at distances $R_{form} \sim 0.6$fm.

Now we give two sets of  parameters used for our calculations,\\
\underline{case A}:

$$\Lambda_A = 240 MeV, ~~~~~~m_B = 1.5 GeV, $$

$$
\alpha_B(asym) = 0.240 ; ~~~e_A = \frac{4}{3}\alpha_B(asym)=0.320;
~~~F^{NP}(R) = Ka^2 = 0.0114 \eqno{(16)}$$
\underline{case B}:

$$ \Lambda_B = 280 MeV, ~~~ m_B = 1.0 GeV$$

$$
\alpha_B(asym) = 0.343 ;~~~e_B = \frac{4}{3}\alpha_B(asym) = 0.457; ~~~
F^{NP}(R)~~ \mbox{is given by (15)} \eqno{(17)}$$
with $\delta = 1$GeV or $T_g = 0.2$fm

Our calculations of the force (the cases A and B)  at $R < 4a$   are
presented in table 1 together with lattice data from \cite{4} and the P
calculations of $F^P(R)$  with $\Lambda_R=$289 MeV, which  slightly differs
from the value $\Lambda_{\overline{MS}} = 256 (20)$MeV or $\Lambda_R =
$268MeV which was predicted in \cite{4}. We have used $\Lambda_R = $289 MeV
because for this value we get very good $\chi^2  (
\frac{\chi^2}{N_{d.o.f.}} = \frac{5.2}{8})$  whereas for $\Lambda_R =
268$MeV $\frac{\chi^2}{N_{d.o.f.}}=\frac{20.0}{8}$ is too large.
At $R < 4a$ our calculations of $F(R)$  give also a good description of
lattice data: in case A we get  $\frac{\chi^2}{N_{d.o.f.}}=\frac{7.0}{8}$
and in case B $\frac{\chi^2}{N_{d.o.f.}}=\frac{4.5}{8}$ is better (see table
1).

At medium and large separations, $R \geq 4a$, the values of $F(R)$ (the
cases A and B)  are presented in table 2 together with 15 values found on
the lattice. In both cases we have obtained a good agreement with lattice
data. In case A we get $\frac{\chi^2}{N_{d.o.f.}}=\frac{11.6}{15}$ and in
case B -- $\frac{\chi^2}{N_{d.o.f.}}=\frac{16.0}{15}$ which should be
compared with $\frac{\chi^2}{N_{d.o.f.}}=\frac{12.5}{15}$  obtained in
\cite{4} for the parametrization (5)  corresponding Cornell potential.

Thus we can conclude that with running c.c. of background perturbation
theory  it is possible to provide a good description of lattice
static interaction over all interval of separations $R$, $a \leq R \leq 23a$
or $0.05$fm$\la R \leq 1.1.$fm. It holds in both cases, A and B. The
c.c. $\alpha_B(R)$ strongly depends on the parameters $m_B$ and $\Lambda_R$
(mostly on $( \frac{m_B}{\Lambda_R})^2$) but
still we are not able to distinguish between two sets of these parameters.
There are two reasons for it. First, the  lattice errors are still too large
in \cite{4,5} and, secondly, the fitted values of $m_B$  and $\Lambda_R$
depend indirectly on NP contribution to the force. At present this
$F^{NP}(R)$ is not known  unambiguously.
The constant force, $F^{NP}(R)=Ka^2$, which is widely
used in lattice analysis, corresponds to vanishing $T_g~(T_g = 0)$  whereas
the finite value of $T_g=0.2$fm was obtained in \cite{12}.
So it would be important to determine
$F^{NP}(R)$
from  the lattice  measurements in
unambiguous way.
Then, having more precise lattice data on the force and knowing $T_g$  with
good accuracy, one has an opportunity to determine unambiguously $m_B$  and
$\Lambda_R$.

This $\alpha_B(R)$  is an analytical function of $R$ at
any distances and freezing at large separations.
The main difference
between cases A and B considered here is in
the asymptotic value of
$\alpha_B(R)$ or in constant $e$: $e_A = 0.32$ and $e_B \cong 0.46$, i.e. 
$e_B$ is about 40\% larger than $e_A$. It is interesting to note  that our 
$e_A$ is very close to the fitted values of $e = 0.31 \div 0.315$  obtained
in \cite{5}  for static potential with linear confining term which
corresponds to our case A.
 On the other hand our value of $e_B \cong 0.46$  is close to Coulomb
 parameter $e_{phen}$, usually used in phenomenological potential models,
 e.g. to describe heavy quarkonia spectra the values of
 $e_{phen}(n_f=3)\sim 0.45 \div 0.55$  are needed. Notice also that the
 fitted values of $e_A$ and $e_B$ found here are about $(30 \div 60)\%$
 larger than the value of $e$(string)$=\frac{\pi}{12}\cong 0.262$ predicted
 in bosonic string theory at large distances \cite{2}.

Author is grateful to Prof. Yu.A.Simonov for  fruitful discussions. This
work was partly supported by RFFI-DFG  grant 96-02-00088G.

\newpage

{\bf Table 1.}~Lattice data and theoretical calculations
of the  force $F(R)$

~~~~~~~~~~~~~at small distances, $R < 4a$.


\vspace{1cm}
\begin{tabular}{|c|c|c|c|c|}\hline
$R/a$ & PT & Case A & Case B & Lattice data \\
    & $\Lambda_R  =289$MeV$^{a)}$ & $\Lambda_R  =240$MeV$^{b)}$ &
    $\Lambda_R  =280$MeV$^{c)}$ & from \cite{4}$^{d)}$\\ \hline
&&&& \\
1.2071 & 0.1662 & 0.1623 & 0.1639 & 0.1607 (55) \\ \hline
&&&& \\
1.7071 & 0.0956 & 0.0944 & 0.0940 & 0.0930(23) \\ \hline
&&&& \\
2.1180 & 0.0686 & 0.0685 & 0.0674 & 0.0664(48) \\ \hline
&&&& \\
2.5322 &   0.0527 & 0.0530 & 0.0517 & 0.0523(6) \\ \hline
&&&& \\
2.9142 & 0.0432 & 0.0438 & 0.0424 & 0.0424(57) \\ \hline
&&&& \\
3.0811 & 0.0400 & 0.0404 & 0.0393 & 0.0368(49)\\ \hline
&&&& \\
3.3839 & 0.0353 & 0.0361 & 0.0348 & 0.0371(14) \\ \hline
&&&& \\
3.9241 & 0.0292 & 0.0302 & 0.0287 & 0.0290(6) \\ \hline
\end{tabular}

\bigskip
a) calculations in perturbative theory with $\Lambda_R=289$Mev;

b) $m_B=1.5$GeV, $Ka^2=0.0114$;

c) $m_B=1.0$GeV and $F^{NP}(R)$ is given by (15) with $Ka^2=0.0114$,
$\delta=1$GeV;

d) lattice errors are found from the paper \cite{4}, Table 3
($\beta=6.5;~a^{-1}=4,13$GeV).

\newpage

\hspace{3cm} {\bf Table 2.}~ Lattice data and theoretical calculations of

\hspace{5cm} force $F(R)$ at large distances, $R \geq 4a$.

\begin{center}
\bigskip

\begin{tabular}{|l|c|c|c|} \hline
~~$R/a$  & Case A$^{a)}$ & Case $B^{b)}$ & Lattice data$^{c)}$ \\
&$\Lambda_R = 240$MeV & $\Lambda_R=280$MeV & from \cite{4}\\ \hline
&&& \\
4.2361 & 0.0278 & 0.0268 & 0.0286(8) \\ \hline
&&& \\
5.0645 & 0.0231 & 0.0218 & 0.0226(5) \\ \hline
&&& \\
5.8284 & 0.0204 & 0.0191 & 0.0196(19)\\ \hline
&&& \\
6.1623 & 0.0194 & 0.0183 & 0.0190(16)\\ \hline
&&& \\
6.7678 & 0.0181 & 0.0171 & 0.0174(9) \\ \hline
&&& \\
7.6056 & 0.0167 & 0.0159 & 0.0161(10) \\ \hline
&&& \\
8.2426 & 0.0160 & 0.0152 & 0.0161(11) \\ \hline
&&& \\
~9.00 & 0.0152 & 0.0146 & 0.0149(3)\\ \hline
&&& \\
11.00 & 0.0139 & 0.0136 & 0.0130(4)\\ \hline
&&& \\
13.00 & 0.0133 & 0.0131 & 0.0134(5)\\ \hline
&&& \\
15.00 & 0.0128 & 0.0127 & 0.0121(6)\\ \hline
&&& \\
17.00 &  0.0125 & 0.0126 & 0.0132(6)\\ \hline
&&& \\
19.00 & 0.0123 & 0.0124 & 0.0117(6)\\ \hline
&&& \\
21.00 & 0.0121 & 0.123 & 0.0125(8)\\ \hline
&&& \\
23.00 & 0.0120 & 0.0122 & 0.0126(7)\\ \hline
\end{tabular}

\end{center}

\bigskip

\hspace{2.5cm} a) see footnote (b) in Table 1;

\hspace{2.5cm} b) see footnote  (c)  in Table 1;

\hspace{2.5cm} c) lattice data for $a^2\frac{\Delta V}{\Delta R}$  are taken
from \cite{4}, Table 1.

\end{document}